\documentclass[pre,twocolumn,superscriptaddress]{revtex4}
\usepackage{color}
\usepackage{epsfig}
\usepackage{subfigure}
\usepackage{amsmath}
\usepackage{amsfonts}

\begin{document}

\title[Alternative Derivation of the Partition Function for Generalized Ensembles]{Alternative Derivation of the Partition Function for Generalized Ensembles}

\author{Jonathan L. Belof\footnote{Present address:\\Lawrence Livermore National Laboratory\\7000 East Avenue., Livermore, CA 94550} and Brian Space}
\address{Department of Chemistry, University of South Florida\\4202 E. Fowler Ave., Tampa, FL 33620\\}

\begin{abstract}
A pedagogical approach for deriving the statistical mechanical
partition function, in a manner that emphasizes the key role of entropy in
connecting the microscopic states to thermodynamics, is introduced.
The connections between the combinatoric formula $S= k \ln
W$ applied to the Gibbs construction, the Gibbs entropy, $S = -k
\sum\limits_i p_i \ln p_i$, and the microcanonical entropy expression
$S= k \ln \Omega$ are clarified. The condition for microcanonical
equilibrium, and the associated role of the entropy in the thermodynamic potential is
shown to arise naturally from the postulate of equal {\itshape a
priori} states.  The derivation of the canonical partition function
follows simply by invoking the Gibbs ensemble construction at constant
temperature and using the first and second law of thermodynamics
(\emph{via} the fundamental equation $dE = TdS - PdV + \mu dN$) that
incorporate the conditions of conservation of energy and composition
without the needs for explicit constraints; other ensemble follow easily.
The central role of the entropy in establishing equilibrium for a given
ensemble emerges naturally from the current approach.  Connections to
generalized ensemble theory also arise and are presented in this context.

\end{abstract}

\maketitle

\section{Introduction}

In deriving the partition function for a desired ensemble, the most
common approach is to maximize an entropy function with constraints
appropriate to the thermodynamic condition. While equivalent to the
approach proposed below, such a method (called the traditional
approach hereafter) does not make clear to students the explicit role
of the assumption of equal {\itshape a priori} states and the
corresponding role of the entropy in the thermodynamic potential for
the microcanonical ensemble. Indeed, $S= k \ln \Omega$ is often taken
as a postulate\cite{ross} and its connection to the statistical
formula $S= k \ln W$ (appearing on Boltzmann's tombstone) is not
obvious. Further, in the traditional approach, the role of the entropy
in understanding equilibrium in non-isolated, open ensembles can be
confusing.  We note in passing that concerns over the rigor
of the method of most probable distribution prompted Darwin and Fowler
to develop a derivation of the partition function based upon
complex analysis.\cite{huang}

Also, infrequently stressed is the Gibbs entropy, $S = -k
\sum\limits_i p_i \ln p_i$, where $p_i$ is the probability of finding
a system in a given state, which can be invoked for any equilibrium
ensemble and associated state probabilities.\cite{ejppaper} It is a
direct consequence of the statistical entropy formula, $S= k \ln W$,
in conjunction with the Gibbs construction of an ensemble that
contains a large number of macroscopic subsystems, each consistent
with the desired thermodynamic variables; $W$ gives the number of
possible realizations within the Gibbs construction for the ensemble
under consideration.  The Gibbs entropy also permits the derivation of
the connection between the characteristic thermodynamic function and
the partition function for a given ensemble without further appeal to
thermodynamic expressions, as is required in the traditional approach.

In the present approach, first, the connections between the
statistical formula $S= k \ln W $, the Gibbs entropy, $S = -k
\sum\limits_i p_i \ln p_i$, and the microcanonical entropy expression
$S= k \ln \Omega$ are clarified. The condition for microcanonical
equilibrium, and the associated role of the entropy in the
thermodynamic potential then arises from the postulate of equal
{\itshape a priori} states.  The derivation of the canonical partition
function follows by invoking the Gibbs construction and the first and
second law of thermodynamics \emph{via} the fundamental equation, $dE
= TdS - PdV + \mu dN$, that incorporates the conditions of
conservation of energy and composition without the needs for explicit
constraints. The role of the temperature (coming from the constraint
of total energy and an appeal to appropriate thermodynamic
relationships in the traditional approach) is immediately apparent and
also introduced \emph{via} the fundamental equation.  The need for
explicit maximization of any function is thus also avoided. Legendre
transforming a particular thermodynamic function to include desired
thermodynamic control variables for an ensemble of interest and
invoking equilibrium leads to the corresponding partition
function. Using the resulting probabilities in the Gibbs entropy
expression directly connects the partition function to the
thermodynamic potential. The central role of the entropy in
establishing equilibrium for a given ensemble emerges naturally from
the current approach. Connections to generalized ensemble theory also
arise and are presented in this context.

The present approach is novel in providing clarity as to the roles
played by the different formulas and physical quantities of interest.
Further, it makes explicit the assumptions inherent in deriving the
partition function for an ensemble and provides its direct connection
to the relevant thermodynamic potential in a systematic fashion. This
approach also makes deriving the partition function for a given
ensemble a simplified, straight-forward process, even for more
challenging examples such as the isothermal-isobaric ensemble. Using
this approach in the classroom has led to better retention and
understanding of the foundations of statistical mechanics and an
ability for students to confidently apply the machinery to problems
that arise in their subsequent work.

\section{The Gibbs Entropy and the Microcanonical Ensemble}
\label{sec:boltzman_law}

We begin by introducing the concept of an ensemble of replicas that
describe the molecular states corresponding to a given macrostate;
this picture is referred to as the ``Gibbs construction" herein, due
to it's original introduction by Gibbs\cite{gibbs,mcquarrie1} who
addressed many of the subtleties inherent\cite{ross} in the
formulation of statistical mechanics.  Consider a collection of
macroscopic molecular ``subsystems" of $N$ molecules within a volume
$V$, each of which is part of the larger Gibbs construction, the
totality of which is known as the ``system".  No other constraints
have yet been imposed, \emph{i.e.} the system's macrostate is
otherwise unspecified.  It is desirable to define the microscopic
statistics of this system as thoroughly as possible and then apply any
other constraints at the end.

Let the total number of subsystems in our collection be known as
$\Omega$.  Then let $\omega_i$, the occupation number, denote the
number of subsystems from this collection that are in the same thermodynamic state.
These occupations will thus take on a large value in the thermodynamic limit
and they obey a sum rule, $\sum\limits_i \omega_i = \Omega$. Note, technically
the energy is course-grained, \emph{i.e.} specified to within a small
but otherwise arbitrary range (these arguments are presented in
detail elsewhere\cite{ross,huang}) and the results are
insensitive to this choice.

First, consider the following combinatoric formula:
\begin{eqnarray}
S_e = k \ln W \left\{ \omega \right\} = k \ln\frac{\Omega!}{\omega_1! \omega_2! ... }
\end{eqnarray}
$W\left\{ \omega \right\}$ is the number of ways in which the set of
occupations $\left\{ \omega \right\}$ may be arranged consistent with
the given macrostate.  First, it is to be shown that when evaluated at
fixed energy, this quantity $S_e$ may be identified with the
thermodynamic entropy of the ensemble of systems at equilibrium, with
each systems entropy given by S = $\frac{S_e}{\Omega}$.
Note, the expression necessarily involves the logarithm of the
combinatoric expression to make the entropy an extensive property;
for two independent systems the possible number of
arrangements is the product of those for the individual systems, $S
= k \ln \left\{W_1 W_2\right\} = k \ln \left\{W_1\right\} + k \ln \left\{W_1\right\} = S_1 +S_2$.

Next the connection between the combinatoric formula $S_e=k \ln W$ and
the Gibbs entropy is presented; the details of this have been given
elsewhere.\cite{mcquarrie2} Applying the Sterling
approximation\cite{eyring} to the factorial function gives the entropy:
\begin{eqnarray}
S_e = k \left\{ \Omega \ln \Omega - \Omega - \sum_i \omega_i \ln \omega_i + \sum_i \omega_i \right\} \nonumber \\
\end{eqnarray}
Further substituting $\omega_i = \Omega p_i$ where $p_i = { \omega_i /
\Omega}$ is identified as the probability of a particular state and
using the property of the natural log gives a system's entropy as:\cite{mcquarrie2}
\begin{eqnarray}
{S}= -k \sum_i p_i \ln p_i \label{eq:ge}
\end{eqnarray}

This is the Gibbs entropy in an as yet unspecified ensemble with its
associated probabilities; the Gibbs entropy is an entirely general
definition that, for any equilibrium ensemble, specifies the
relationship between the partition function and the associated
characteristic thermodynamic function.

Now, specializing to a set of microcanonical subsystems, and
invoking the equilibrium principle of equal \emph{a priori} states,
\emph{i.e.} $p_i = {1 / \Omega}$, gives the well known result:
\begin{eqnarray}
S = -k \sum_i^\Omega \frac{1}{\Omega} \ln \frac{1}{\Omega} = k \ln
\Omega
\end{eqnarray}
It is also simple and useful to show that the Gibbs entropy, and thus
the thermodynamic entropy, is maximized
microcanonically\cite{chandler} by the state-independent probabilities
$p = p_i = 1/ \Omega$. Proceeding, taking the derivative of Equation
\ref{eq:ge} and setting it to zero as
\begin{eqnarray}
\frac{\partial}{\partial p_j} \left(-k \sum_i p_i \ln p_i\right) = 0
\end{eqnarray}
gives $p_j = 1 / e$, a constant value independent of
the summation index. Thus, normalizing the probabilities as,
$\sum\limits_i^\Omega p_i = 1$ immediately yields $p_i =  1 / \Omega$.

Thus, for an isolated system, the assumption of equal \emph{a priori}
states leads to a probability $p_i$ that is independent of index,
\emph{i.e.} every subsystem has energy $E$ by construction. Further,
the characteristic maximum entropy in the microcanonical equilibrium
ensemble also follows.  Then applying the Gibbs entropy expression leads
to the identification of the thermodynamic entropy as the
characteristic function of the microcanonical ensemble and gives its relationship to
the $N,V,E$ partition function, $\Omega(E)$, which can also be
interpreted as the density of states\cite{ross} at that energy.

\section{A Simplified Derivation of the Canonical Partition Function}
\label{sec:canonical}

Specializing the Gibbs construction from the previous section to 
include temperature, we have a collection of subsystems all possessing the same $N,V,T$.
This can be thought of by placing the subsystems in contact
with a large heat bath of temperature $T$.\cite{chandler,ross}
We now imagine that each subsystem (after having achieved equilibrium with
the heat bath by definition) is to be insulated and the energy of the $i^{th}$
subsystem is measured as $E_i$, and for which there is also an associated
macroscopic entropy $S_i$.  Of great importance, we also note that the
thermodynamic energy $E_i$ is exactly equal to the microscopic configurational
energy of the subsystem upon insulation.  Furthermore, the details and/or rates involved
in the insulation process are irrelevant for an equilibrium ensemble.

Using the earlier result, the entropy for a collection of subsystems with
a specified energy $E_i$ is $S_i = k \ln \Omega(E_i)$, where $\Omega$ is the
number of subsystems with energy $E_i$ in the ensemble.

Consider the ratio of the density at energies $E_{i+1} > E_i$:

\begin{eqnarray}
\frac{\Omega(E_{i})}{\Omega(E_{i+1})} = \frac{e^{\frac{1}{k}S_{i}}}{e^{\frac{1}{k}S_{i+1}}} = e^{-\frac{1}{k}(S_{i+1} - S_{i})} \label{eq:ratio}
\end{eqnarray}

The fundamental equation of thermodynamics\cite{alberty} is now invoked:

\begin{eqnarray}
dE(S,V,N) = TdS - PdV + \mu dN \label{eq:fe}
\end{eqnarray}

The canonical ensemble is given by a state with well defined
thermodynamic variables, $N, V, T$.  So the energy function,
$E(S,V,N)$ is Legendre transformed to a new thermodynamic
function, the Helmholtz free energy, $A(T,V,N)$ \emph{via}:

\begin{eqnarray}
A = LT\left\{ E \right\} = E - S \frac{\partial E}{\partial S} = E - ST \\
dA = dE - TdS - SdT
\end{eqnarray}
where the condition for canonical equilibrium is that
$dA = 0$ and $N,V, T$ are constant, giving:
\begin{eqnarray}
0 = dE - TdS \\
dS = \frac{1}{T}dE 
\end{eqnarray}

Integrating between two state points gives:

\begin{eqnarray}
\int\limits_i^{i+1} dS = \frac{1}{T} \int\limits_i^{i+1} dE \\
S_{i+1} - S_{i} = \frac{1}{T} \left( E_{i+1} - E_{i} \right) \label{eq:ds}
\end{eqnarray}

Note, the constraints of fixed temperature, particle number and volume
have been explicitly enforced by using the fundamental equation,
Equation \ref{eq:fe} and dropping differential terms that are fixed
canonically.

Substituting Equation \ref{eq:ds} into Equation \ref{eq:ratio} gives:
\begin{eqnarray}
\frac{\Omega(E_{i})}{\Omega(E_{i+1})} = e^{-\frac{1}{k}(S_{i+1} - S_{i})} = \frac{e^{\beta E_i}}{e^{\beta E_{i+1}}} \label{eq:om} \\
\frac{p_i}{p_{i+1}} = \frac{e^{-\beta E_i}/Q}{e^{-\beta E_{i+1}}/Q}
\end{eqnarray}
where $\beta = 1/kT$ and $p_i = \frac{1/\Omega_i}{\sum\limits_i 1/\Omega_i}$
(\emph{i.e.} the probability of choosing the $i^{th}$ state from the entire
ensemble at equilibrium with the heat bath).
The normalization factor, $Q = \sum\limits_i e^{-\beta E_i}$, may be readily recognized as the
canonical partition function.  Most importantly, we note that the
insulation procedure applied to each subsystem has allowed us to
identify the macroscopic energy (and entropy) of that microcanonical system,
with the microscopic energy of the molecular configuration present at the time
of insulation.

We can now proceed to use the Gibbs entropy expression, Equation \ref{eq:ge}, substituting
the canonical expression for $p_i$ to obtain:
\begin{eqnarray}
\label{eq:cpf}
\nonumber
S = - k \sum_i { e^{-\beta E_{i}} \over Q} \ln { e^{-\beta E_{i}} \over Q} \\
\nonumber
S = k {\ln Q \over Q} \sum_i e^{-\beta E_{i}}  + k \beta \sum_i {E_i e^{-\beta E_{i}} \over Q }\\
TS = kT \ln Q + \langle E \rangle_{NVT} 
\end{eqnarray}

Above, $\langle E \rangle_{NVT}$ represents the canonical average energy that is
identified with the thermodynamic energy, $E$.\cite{mcquarrie1,ross}
Thus, the relationship $A = E - TS = -kT \ln Q$ is obtained directly
from the Gibbs entropy. Note, the Gibbs entropy is defined for any set
of probabilities and, as was shown above, is simply a consequence of
the combinatoric formula $S=k \ln W$ interpreted in the context of
the Gibbs construction.  Thus, an entropy can be associated even with
nonequilibrium probabilities. However, in that case, the entropy does
not play the role of being the constrained maximized quantity that it 
does at equilibrium and its utility, in such circumstances, is unclear.

Further note, the role of temperature is
introduced \emph{via} the fundamental equation without further appeal
to thermodynamic relationships.  This emphasizes the role of
temperature as the system is in contact with a heat bath -- different
energy ranges are now accessible with canonical
probabilities. The ability of a diathermal system to exchange energy
with its surroundings also clarifies how the concepts of work and
entropy make sense for an open system and provides their relationship
to the temperature.

\begin{figure}[htp]
\begin{center}
\includegraphics[width=3.3 in]{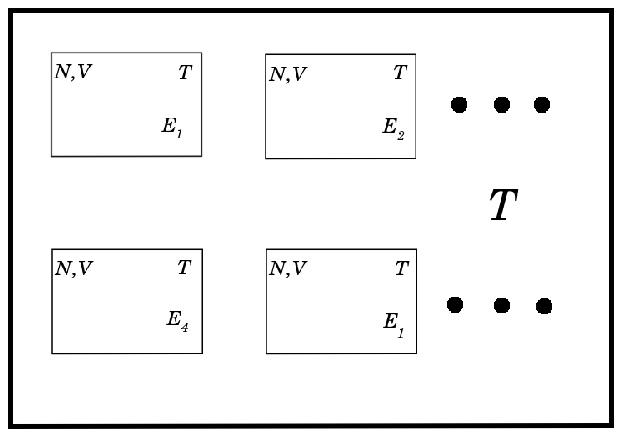}
\caption{\label{fig:canonical_ensemble}Gibbs construction for the canonical ensemble.  The subsystems of identical $N,V$ are in thermal equilibrium with a large bath at temperature $T$.}
\end{center}
\end{figure}

\section{Grand Canonical Partition Function}
\label{sec:grand_canonical}

Using the Gibbs construction from the previously derived canonical ensemble, the constraint that all subsystems possess identical $N$ can now be relaxed.  We now consider the ratio of subsystem, having chosen a particular system from the energy level $E_i$ with $N_j$ molecules:

\begin{eqnarray}
\frac{\Omega(N_j, E_i)}{\Omega(N_{j+1}, E_{i+1})} = e^{-\frac{1}{k}(S_{i+1,j+1} - S_{i,j})}
\end{eqnarray}

As before, the fundamental equation of thermodynamics can be relied upon to relate the change in entropy to the other variables of our ensemble.  In this case, the probabilities that will generate the macrostate corresponding to constant $\mu, V, T$ are desired.  After doing so, the entropy \emph{via the Boltzmann law} is used to determine the microscopic states.

Legendre transforming the canonical thermodynamic equation to substitute $\mu$ for $N$:
\begin{eqnarray}
J = LT\left\{ A \right\} = A - N\frac{\partial A}{\partial N} \nonumber \\
= A - \mu N \\
dJ = dA - Nd\mu - \mu dN \nonumber \\
= dE - TdS - SdT - Nd\mu - \mu dN
\end{eqnarray}
where at equilibrium $dJ = 0$ and with constant $\mu, V, T$:
\begin{eqnarray}
0 = dE - TdS - \mu dN \\
dS(\mu, V, T) = \frac{1}{T}\left(dE - \mu dN\right)
\end{eqnarray}

Upon integrating, the difference equation for the entropy is found:

\begin{eqnarray}
\int\limits^{i+1, j+1}_{i, j} dS = \frac{1}{T}\left( \int\limits^{i+1}_{i} dE - \mu \int\limits^{j+1}_{j} dN \right) \\
S_{i+1,j+1} - S_{i,j} = \frac{1}{T}\left[ E_{i+1} - E_i - \mu \left(N_{j+1} - N_j\right)\right]
\end{eqnarray}

The ratio of microstates then becomes:

\begin{eqnarray}
\frac{\Omega(N_j, E_i)}{\Omega(N_{j+1}, E_{i+1})} = e^{-\frac{1}{k}(S_{i+1,j+1} - S_{i,j})} \nonumber \\
= e^{-\frac{1}{kT}(E_{i+1} - E_{i} - \mu N_{j+1} + \mu N_j)} \nonumber \\
= \frac{e^{-\beta \mu N_j} e^{\beta E_i}}{e^{-\beta \mu N_{j+1}} e^{\beta E_{i+1}}}
\end{eqnarray}
and since $p_{j,i} = \frac{1/\Omega_{j,i}}{\sum\limits_j \sum\limits_i 1/\Omega_{j,i}}$
\begin{eqnarray}
\frac{p_{j,i}}{p_{{j+1},{i+1}}} = \frac{e^{\beta \mu N_j}e^{-\beta E_i}/\Xi}{e^{\beta \mu N_{j+1}}e^{-\beta E_{i+1}}/\Xi}
\end{eqnarray}
where the normalization factor $\Xi$ is the \emph{grand canonical
partition function}:

\begin{eqnarray}
\Xi(\mu, V, T) = \sum\limits_j e^{\beta \mu N_j} \sum\limits_i e^{-\beta E_i}
\end{eqnarray}

It may be noted that in this case the constraint
of constant $N$ was merely relaxed, Legendre transformed to the
corresponding macrostate, and the partition function then followed
quite naturally and simply. The resulting probabilities can now be
substituted into the Gibbs entropy and the relationship between the
thermodynamic potential and partition function is thus directly
established.

\begin{figure}[htp]
\begin{center}
\includegraphics[width=3.3 in]{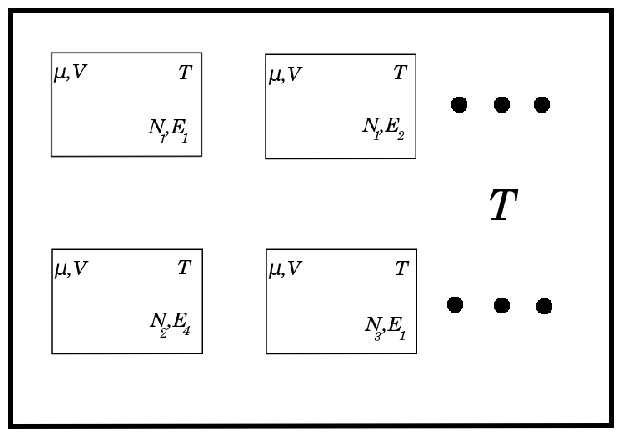}
\caption{\label{fig:grand_canonical_ensemble}Gibbs construction for
the grand canonical ensemble.  $\mu$ has been Legendre transformed to
replace $N$ as the macroscopic constant, and so the set of subsystems
includes those of differing $N$ values.}  
\end{center}
\end{figure}

\section{Isothermal-Isobaric Partition Function}
\label{sec:isobaric}

Having successfully derived the grand canonical partition function by relaxing the constant $N$ constraint, it can now be shown that the isothermal-isobaric ensemble is generated by starting with constant $NVT$ and relaxing the condition of constant $V$.  Consider the number of subsystems with both volume $V_i$ and energy $E_j$:

\begin{eqnarray}
\frac{\Omega(V_j, E_i)}{\Omega(V_{j+1}, E_{i+1})} = e^{-\frac{1}{k}(S_{i+1,j+1} - S_{i,j})}
\end{eqnarray}

Legendre transforming our desired variables to a new characteristic function $G$, and then applying the condition of equilibrium $dG = 0$ and our constant differential terms:
\begin{eqnarray}
G = A - V\frac{\partial A}{\partial V} = A + PV \\
dG = dA + PdV + VdP = 0 \nonumber \\
= dE - TdS - SdT + PdV + VdP \\
\rightarrow dS = \frac{1}{T}\left(dE + PdV\right)
\end{eqnarray}

and so the ratio of observable subsystems becomes:

\begin{eqnarray}
\frac{\Omega(V_j, E_i)}{\Omega(V_{j+1}, E_{i+1})} = e^{-\frac{1}{k}(S_{i+1,j+1} - S_{i,j})} \nonumber \\
= e^{-\frac{1}{kT}[E_{i+1} - E_i + P(V_{i+1} - V_i)]} \nonumber \\
= \frac{e^{\beta E_i}e^{\beta PV_j}}{e^{\beta E_{i+1}}e^{\beta PV_{j+1}}}
\end{eqnarray}
and since $p_{j,i} = \frac{1/\Omega_{j,i}}{\sum\limits_j \sum\limits_i 1/\Omega_{j,i}}$
\begin{eqnarray}
\frac{p_{i,j}}{p_{i+1,j+1}} = \frac{e^{-\beta E_i}e^{-\beta PV_j} / \Delta}{e^{-\beta E_{i+1}}e^{-\beta PV_{j+1}} / \Delta} 
\end{eqnarray}
where the normalization factor,
\begin{eqnarray}
\Delta = \sum_j e^{-\beta PV_j} \sum_i e^{-\beta E_i}
\end{eqnarray}
is the \emph{isothermal-isobaric partition function}.

Again, the associated probabilities can now be substituted into the
Gibbs entropy and the relationship between the thermodynamic potential
and partition function is thus directly established.

\begin{figure}[htp]
\begin{center}
\includegraphics[width=3.3 in]{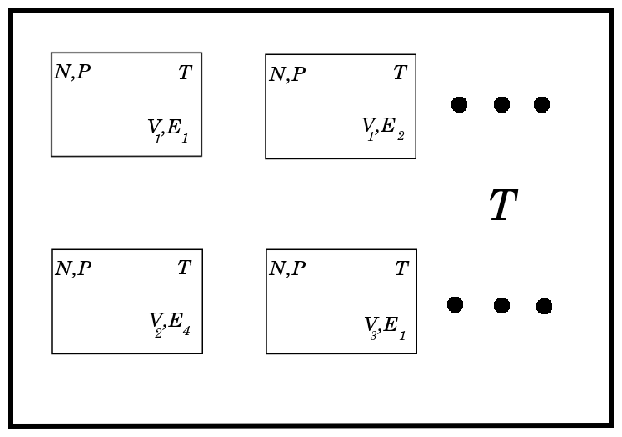}
\caption{\label{fig:isothermal_isobaric_ensemble}Gibbs construction for the isothermal-isobaric ensemble.  $P$ has been Legendre transformed to replace $V$ as the macroscopic constant, and so the set of subsystems includes those of differing $V$ values.}
\end{center}
\end{figure}

\section{Connection with Generalized Ensemble Theory}
\label{sec:gen_ensemble_theory}

Based on the previous sections, it may be noted that if a partition
function for \emph{any} macrostate in thermal equilibrium is to be
derived, one shall \emph{always} arrive at an expression that involves
an exponential function; this follows as a consequence of the
Boltzmann Law.  Furthermore, the partition function being the
normalization factor to express this as a probability means that one
will always have a discrete sum.  Therefore, our partition functions
will always be some variation on a theme amounting to a ``sum of
exponentials."

In the continuum limit, it can be shown that the canonical partition function $Q$ (written as a sum over energy levels) transforms as:

\begin{eqnarray}
Q(N,V,\beta) = \sum_i e^{-\beta E_i} \Omega(E_i) \nonumber \\
\rightarrow \int_0^\infty dE \, e^{-\beta E} \Omega(N,V,E)
\end{eqnarray}
or, in other words, the canonical partition function is the Laplace transform of the microcanonical partition function $\Omega$.

How does it work the other way?  Let's apply the inverse Laplace transform to $Q$:

\begin{eqnarray}
\Omega(N,V,E) = \frac{1}{2\pi i} \oint d\beta\, e^{\beta E} Q(N,V,\beta) \nonumber \\
= \int\limits_{-\infty}^{\infty} d\Omega\, \frac{1}{2\pi i} \int\limits_{\gamma - i\infty}^{\gamma + i\infty} d\beta\, e^{\beta(E - H)}
\end{eqnarray}
where the phase space differential form $d\Omega = (h^{3N} N!)^{-1} dx_1...dx_{3N} dp_1...dp_{3N}$.
Now, $\beta = \sigma + i\tau$ and because no singularity is present in the right-half of the complex plane, the contour may be taken vertically through $\gamma = 0$.  Since $\rm Re(\beta) = 0$ along the integration, the substitution $\beta = -i\tau$ can be made:

\begin{eqnarray} 
\Omega(N,V,E) = \int\limits_{-\infty}^{\infty} d\Omega\, \frac{1}{2\pi} \int\limits_{-\infty}^{\infty} d\tau\, e^{i\tau(E - H)} \nonumber \\
= \int\limits_{-\infty}^{\infty} d\Omega\, \delta(H - E) \label{eq:delta_function}
\end{eqnarray}
where indeed Equation \ref{eq:delta_function} can be identified as the microcanonical partition function.  Thus, any constant energy shell ensemble may be Laplace transformed to an ensemble of a new intensive variable.  As the partition functions are related to one another through the Laplace transform, this is isomorphic to the thermodynamic potentials (to each of which may be associated a particular partition function) being related through the Legendre transform.\cite{zia}

The quantum harmonic oscillator is an illustrative example of how the canonical partition function may be transformed to the microcanonical case:

\begin{eqnarray}
\Omega_{HO} = \frac{1}{2\pi i} \oint d\beta\, e^{\beta E} Q_{HO} \nonumber \\
= \frac{1}{2\pi} \int\limits_{-\infty}^{\infty} d\tau\, e^{-i\tau E} \frac{e^{\frac{1}{2}i\tau \hbar \omega}}{1 - e^{i \tau \hbar \omega} } \nonumber \\
= \frac{1}{2\pi} \int\limits_{-\infty}^{\infty} d\tau\, e^{i\tau\left(\frac{1}{2} \hbar \omega \right) - E} \sum_n e^{i\tau \hbar \omega n} \nonumber \\
= \sum_n \delta \left[\hbar \omega \left( n + \frac{1}{2}  \right) - E\right] \label{eq:nve_qho}
\end{eqnarray}

The reason that the aforementioned ``recipe" for generating the partition function (as outlined in Sections \ref{sec:canonical},\ref{sec:grand_canonical} and \ref{sec:isobaric}) in an arbitrary ensemble works is due to the thermodynamic relations and the Boltzmann law.  The underlying mathematical structure that allows this has also been previously formulated as generalized ensemble theory.\cite{sack,guggenheim,graben,haile,tiller}

\section{Conclusions}
\label{sec:conclusions}

An approach is presented for deriving partition functions that is
an alternative to more common methods. It emphasizes the central role
that (maximizing) the Boltzmann entropy plays in connecting the
molecular states of the system to the observable thermodynamics. Using
this technique in a classroom setting for a beginning graduate class
in statistical mechanics has led to systemization and demystification
of the derivation for useful ensembles. Also, the role of Legendre
transforms to introduce thermodynamic control variables appears
naturally and is tied directly to both the derivation of the ensemble and
corresponding partition function. Within this formalism, students are
clear on how the thermodynamic potential relates to a given ensemble and
the role of equal {\itshape a priori} states. Further,
relating the partition function to the thermodynamic potential using
the Gibbs entropy is straightforward and no further appeal to
thermodynamic expressions is required as the relevant thermodynamic
connection was included from the start of the derivation.

Finally, the similarities between the derivation method demonstrated
and the relations known from generalized ensemble theory have been
noted.  It is our hope that the formulaic approach presented here will
be of utility in both research and pedagogy.

\section{Acknowledgements}
\label{sec:Acknowledgements}

The authors acknowledge funding from the U.S. Department of Energy, Basic Energy Sciences (Grant No. DE0GG02-07ER46470).  Lawrence Livermore National Laboratory is operated by Lawrence Livermore National Security, LLC, for the U.S. Department of Energy, National Nuclear Security Administration under Contract DE-AC52-07NA27344.\\


\end{document}